\documentclass[preprint]{elsarticle}

\usepackage[english]{babel}
\usepackage{amsmath}
\usepackage{amssymb}
\usepackage{epsfig}

\begin{document}
\newcommand{\A}{{\mathcal{A}}}
\newcommand{\dA}{\delta{\mathcal{A}}}
\newcommand{\Od}{{\cal O}}
\newcommand{\degree}{^\circ}
\newcommand{\K}{\textrm{K}}
\newcommand{\h}{{\mathcal{H}}}


\title{The electromagnetic dark sector}%

\author{Jose Beltr\'an Jim\'enez and Antonio L. Maroto}

\address{Departamento de  F\'{\i}sica Te\'orica I,
Universidad Complutense de Madrid, 28040 Madrid, Spain.}
\date{\today}

\begin{abstract}
We consider electromagnetic field quantization in an expanding
universe. We find that the covariant (Gupta-Bleuler) method exhibits
certain difficulties when trying to impose the quantum Lorenz
condition on cosmological scales. We thus explore the possibility
of consistently quantizing  without imposing such a condition. In
this case  there are three physical states, which are  the two
transverse polarizations of the massless photon and a new massless
scalar mode coming from the temporal and longitudinal components
of the electromagnetic field. An explicit example in de Sitter
space-time shows that it is still possible  to eliminate the 
negative norm state and to ensure the  positivity of the 
energy  in this theory.  The new state is decoupled from the
conserved electromagnetic currents, but
 is non-conformally coupled to gravity
and therefore can be excited from vacuum fluctuations by the expanding
background.
The cosmological evolution ensures that the new state modifies
Maxwell's equations in a totally negligible way on  sub-Hubble
scales. However, on cosmological scales it
can give rise to a non-negligible energy density
 which could explain in a natural way the present phase
of accelerated expansion of the universe.
\end{abstract}

\maketitle

\section{Introduction}

The fact that the present phase of accelerated expansion
of the universe \cite{acc} finds no natural explanation within known
physics suggests the possibility that General Relativity could
be inappropriate to describe  gravity on cosmological scales.
Nevertheless, despite the many attempts to find modified gravity theories
with late time accelerated solutions \cite{modified,vector},
the models considered so far are generally plagued by classical or
quantum instabilities, fine-tuning problems or local gravity inconsistencies.

However, apart from gravity, there is another long-range
interaction which could become relevant on cosmological scales,
which is nothing but electromagnetism. It is generally assumed
that due to the strict electric neutrality of the universe on
large scales, the only relevant interaction in cosmology is
gravitation. However, this assumption seems to conflict with
observations which indicate that relatively strong magnetic fields
with very large coherence lengths are present in galaxies and
galaxy clusters, and whose origin is still unknown \cite{Grasso}.

In this work, we consider  (quantum) electromagnetic fields in an 
expanding universe. Unlike previous works, we will not follow the
standard Coulomb gauge approach, but instead we will concentrate
on the covariant quantization method. This 
method is found to exhibit certain difficulties when trying to impose the quantum Lorenz
condition on cosmological scales \cite{Parker}. In order to overcome these problems, 
and instead of introducing ghosts fields,  we explore the possibility of
quantizing without imposing any subsidiary condition. This implies that the
electromagnetic field would contain an additional (scalar)
polarization, similar to that generated in the massless limit
of massive electrodynamics \cite{Deser}. Interestingly, the energy density of quantum
fluctuations of this new electromagnetic state generated during
inflation gets frozen on cosmological scales, giving rise to an 
effective cosmological constant \cite{EM}.

\section{Electromagnetic quantization in flat space-time}

Let us start by briefly reviewing electromagnetic
quantization in Minkowski space-time \cite{Itzykson}.
The action
of the theory reads:
\begin{eqnarray}
S=\int d^4x \left(-\frac{1}{4}F_{\mu\nu}F^{\mu\nu}+A_\mu J^\mu\right)
 \label{action}
\end{eqnarray}
where $J_\mu$ is a conserved current. This action is invariant under
gauge tranformations $A_\mu\rightarrow A_\mu+\partial_\mu \Lambda$
with $\Lambda$ an arbitrary function of space-time coordinates.
At the classical level, this action gives rise to the well-known
Maxwell's equations:
\begin{eqnarray}
\partial_\nu F^{\mu\nu}=J^\mu.
\label{Maxwell}
\end{eqnarray}
However, when trying to quantize the theory, several problems
arise related to the redundancy in the description  
due to the gauge invariance. Thus in particular, we find that it is 
not possible to construct a propagator for the $A_\mu$ field
and  also that "unphysical" polarizations of the 
photon field are present.  Two different approaches are usually followed in order
to avoid these difficulties. In the first one, which is the basis
of the Coulomb gauge quantization, the gauge invariance of the
action (\ref{action}) is used to eliminate the "unphysical"
degrees of freedom. With that purpose the (Lorenz) condition
$\partial_\mu A^\mu=0$ is imposed by means of a suitable gauge
transformation. Thus, the equations of motion reduce to:
\begin{eqnarray}
\Box A_\mu=J_\mu.\label{eqLo}
\end{eqnarray}
The Lorenz condition does not fix completely the gauge freedom,
still it is possible to perform residual gauge transformations
$A_\mu\rightarrow A_\mu+\partial_\mu \theta$, provided $\Box
\theta=0$. Using this residual symmetry and taking into account
the form of equations (\ref{eqLo}), it is possible to eliminate
one additional component of the $A_\mu$ field in the
asymptotically free regions
 (typically $A_0$) which means $\vec \nabla \cdot \vec A=0$,
so that finally the temporal and longitudinal
photons are removed and we are left with the two
transverse polarizations of the massless free photon, which are the
only modes (with positive energies)
which are quantized in this formalism.

The second approach is  the basis of the covariant (Gupta-Bleuler)
and path-integral formalisms. The starting point is a modification of the
action
in (\ref{action}), namely:
\begin{eqnarray}
S=\int d^4x \left(-\frac{1}{4}F_{\mu\nu}F^{\mu\nu}+\frac{\xi}{2}
(\partial_\mu A^\mu)^2+ A_\mu J^\mu\right).
 \label{actionGB}
\end{eqnarray}
This action is no longer invariant under general gauge transformations, but
only under residual ones.
The equations of motion obtained from this action now read:
\begin{eqnarray}
\partial_\nu F^{\mu\nu}+\xi\partial^\mu(\partial_\nu
A^\nu)=J^\mu.
\label{fieldeq}
\end{eqnarray}
In order to recover Maxwell's equation, the Lorenz condition must
be imposed so that the $\xi$ term disappears. At the classical
level this can be achieved by means of appropriate boundary
conditions on the field. Indeed, taking the four-divergence of the
above equation, we find:
\begin{eqnarray}
\Box(\partial_\nu A^\nu)=0
\end{eqnarray}
where we have made use of current conservation. This means that
the field  $\partial_\nu A^\nu$ evolves as a free scalar field, so
that if it vanishes for large $\vert t \vert$, it will vanish at 
all times. At the quantum level, the Lorenz condition cannot be
imposed as an operator identity, but only in the weak sense
$\partial_\nu A^{\nu \,(+)}\vert \phi\rangle=0$, where $(+)$
denotes the positive frequency part of the operator and $\vert
\phi\rangle$ is a physical state. This condition is
equivalent to imposing $[{\bf a}_0(\vec k) +{\bf a}_\parallel(\vec
k)] |\phi\rangle=0$, with ${\bf a}_0$ and ${\bf a}_\parallel$ the
annihilation operators corresponding to temporal and longitudinal
electromagnetic states. Thus, in the covariant formalism, the
physical states contain the same number of temporal and
longitudinal photons, so that their energy densities, having
opposite signs, cancel each other.

Thus we see that also in this case, the Lorenz condition seems to
be essential in order to recover standard Maxwell's equations and
get rid of the negative energy states.

\section{Covariant quantization in an expanding universe}

So far we have only considered Maxwell's theory in flat
space-time, however when we move to a curved background, and in
particular to an expanding universe, then consistently imposing
the Lorenz condition in the covariant formalism turns out to be
difficult to realize. Indeed, let us consider the curved
space-time version of action (\ref{actionGB}):
\begin{eqnarray}
S=\int d^4x
\sqrt{g}\left[-\frac{1}{4}F_{\mu\nu}F^{\mu\nu}+\frac{\xi}{2}
(\nabla_\mu A^\mu)^2+ A_\mu J^\mu\right]
 \label{actionF}
\end{eqnarray}
Now the modified Maxwell's equations read:
\begin{eqnarray}
\nabla_\nu F^{\mu\nu}+\xi\nabla^\mu(\nabla_\nu A^\nu)=J^\mu
\label{EMeqexp}
\end{eqnarray}
and taking again the four divergence, we get:
\begin{eqnarray}
\Box(\nabla_\nu A^\nu)=0\label{minimal}
\end{eqnarray}
We see that once again  $\nabla_\nu A^\nu$  behaves as a scalar
field which is decoupled from the conserved electromagnetic
currents, but it is non-conformally coupled to gravity. This means
that, unlike the flat space-time case,  this field can be excited
from quantum vacuum fluctuations by the expanding background in a
completely analogous way to the inflaton fluctuations during
inflation. Thus this poses the question of the validity of the
Lorenz condition at all times.

In order to illustrate this effect, we will present a toy example.
Let us consider quantization in the absence of currents, in a
spatially flat expanding background, whose metric is written in
conformal time as $ds^2=a(\eta)^2(d\eta^2-d\vec x^2)$ with
$a(\eta)=2+\tanh(\eta/\eta_0)$ where $\eta_0$ is constant. This
metric possess two asymptotically Minkowskian regions in the
remote past and far future. We  solve the coupled system of
equations (\ref{EMeqexp}) for the corresponding Fourier modes
of the conformal field $\A_\mu=(aA_0,\vec A)$,
which are defined as $\A_\mu(\eta,\vec x)= \int d^3k\A_{\mu
\vec k}(\eta) e^{i\vec k \vec x}$. Thus, for a given mode $\vec k$, the
$\A_\mu$ field is  decomposed into temporal,  longitudinal and
transverse components. The corresponding equations read:
\begin{eqnarray}
\A_{0k}''&-&\left[\frac{k^2}{\xi}-2{\mathcal{H}}'
+4{\mathcal{H}}^2\right]\A_{0k}
-2ik\left[\frac{1+\xi}{2\xi}\A_{\parallel k}'
-{\mathcal{H}}\A_{\parallel k}\right]=0 \label{modes}\nonumber\\
\A_{\parallel k}''&-&k^2\xi\A_{\parallel
k}-2ik\xi\left[\frac{1+\xi}{2\xi}\A_{0k}'
+{\mathcal{H}}\A_{0k}\right]=0\nonumber\\
\vec{\A}_{\perp k}''&+&k^2\vec{\A}_{\perp k }=0\label{coupledeq}
\end{eqnarray}
with ${\cal H}=a'/a$ and $k=\vert \vec k\vert$. We see that the transverse 
modes are decoupled
from the background, whereas the temporal and longitudinal ones are
non-trivially coupled to each other and to gravity. Let us prepare our
system  in an initial  state $\vert \phi\rangle$ belonging
to the physical Hilbert space,
i.e. satisfying $\partial_\nu \A^{\nu \,(+)}_{in}\vert \phi\rangle=0$
in the initial flat region.
Because of the expansion
of the universe, the positive frequency modes in the $in$
region with a given temporal or longitudinal polarization $\lambda$
will become a linear superposition of positive and
negative frequency modes in the $out$ region and
with different polarizations $\lambda'$ (we will work in
the Feynman gauge $\xi=-1$), thus we have:
\begin{eqnarray}
\A_{\mu \vec k}^{\lambda \; (in)}=\sum_{\lambda'=0,\parallel}
\left[\alpha_{\lambda\lambda'}(\vec k) \A_{\mu \,\vec k}^{\lambda' \;
(out)}+\beta_{\lambda\lambda'}(\vec k) \overline{\A_{\mu
\,-\vec k}^{\lambda' \; (out)}}\,\right]
\end{eqnarray}
or in terms of creation and annihilation operators:
\begin{equation}
{\bf a}_{\lambda'}^{(out)}(\vec k)=
\sum_{\lambda=0,\parallel}\left[\alpha_{\lambda\lambda'}(\vec k)
{\bf a}_{\lambda}^{(in)}(\vec
k)+\overline{\beta_{\lambda\lambda'}(\vec k)} {\bf
a}_{\lambda}^{(in)\dagger}(-\vec k)\right]
\end{equation}
with $\lambda, \lambda'=0,\parallel$ and
where $\alpha_{\lambda\lambda'}$ and $\beta_{\lambda\lambda'}$
are the so-called Bogoliubov coefficients (see \cite{Birrell} 
for a detailed discussion), which are  normalized in our case according to:
\begin{eqnarray}
\sum_{\rho,\rho'=0,\parallel}(\alpha_{\lambda \rho}\,
\overline{\alpha_{\lambda'\rho'}}\,
\eta_{\rho \rho'}
-\beta_{\lambda \rho}\,\overline{\beta_{\lambda'\rho'}}\,\eta_{\rho\rho'})=
\eta_{\lambda\lambda'}
\end{eqnarray}
with $\eta_{\lambda\lambda'}=diag(-1,1)$ with $\lambda,\lambda'=0,\parallel$.
Notice that the normalization is different from the standard one
\cite{Birrell}, because of the presence of  negative norm states.

\begin{figure}[h]
\begin{center}
{\epsfxsize=8cm\epsfbox{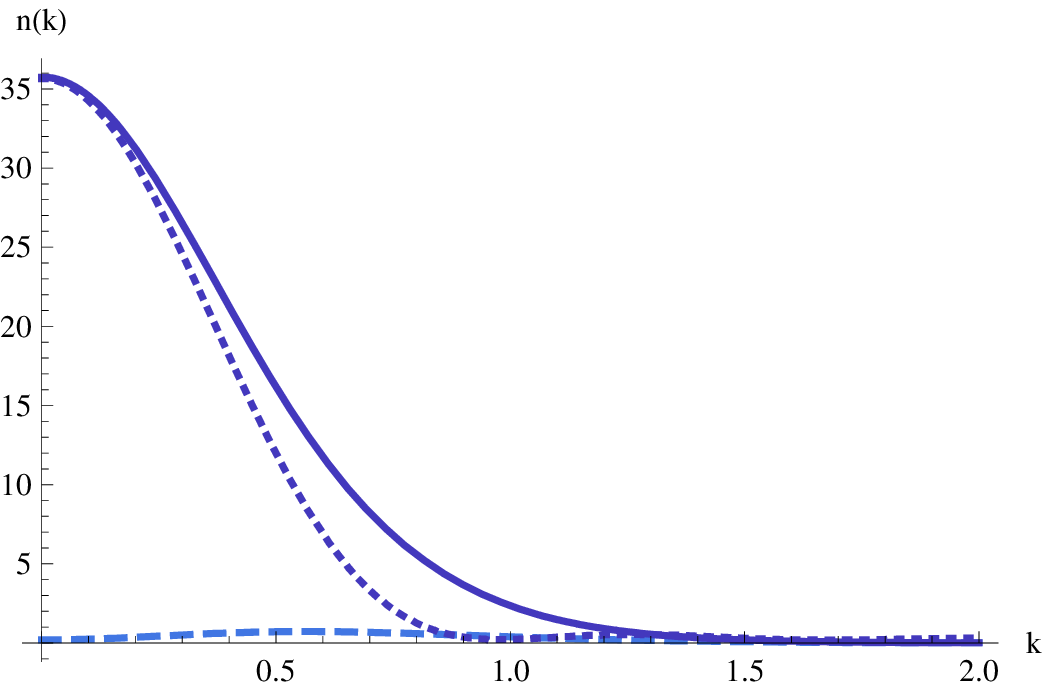}} \end{center}
{\footnotesize {\bf Figure 1:} Occupation numbers for temporal (continuous
line) and longitudinal (dashed line) photons 
in the $out$ region vs.
$k$ in $\eta_0^{-1}$ units. We also show the analytical approximate solution (dotted line) given in the main text.}\label{occupation2}
\end{figure}

Thus, the system will end up in a final state which no longer
satisfies the weak Lorenz condition i.e. in the {\it out} region
$\partial_\nu \A^{\nu \,(+)}_{out}\vert \phi\rangle\neq 0$. This
is shown in Fig. 1, where we have computed  the final number of
temporal and longitudinal photons
$n_{\lambda'}^{out}(k)=\sum_{\lambda}\vert\beta_{\lambda\lambda'}(\vec
k) \vert^2 $, starting from an initial vacuum state with
$n_0^{in}(k)=n_\parallel^{in}(k)=0$. We see that, as commented
above, in the final region $n_0^{out}(k)\neq n_\parallel^{out}(k)$
and the state no longer satisfies the Lorenz condition. Notice
that the failure comes essentially from large scales ($k\eta_0\ll
1$), since on small scales ($k\eta_0\gg 1$), the Lorenz condition
can be restored. This can be easily  interpreted from the fact
that on small scales the geometry can be considered as essentially
Minkowskian. 

We can obtain an approximate analytical solution for the Bogoliubov 
coefficients by assuming an expansion in which $\mathcal{H}$ vanishes  in the {\it in} and {\it out} regions, but it takes a constant value $h$ in the period $(-t_0,t_0)$. Moreover, we shall neglect the production of longitudinal modes (which is justified from our numerical computation). For this simplified problem, the relevant Bogoliubov coefficients are given by:
\begin{align}
\beta_{\parallel 0}=&\frac{1}{2a_{out}^2}e^{2t_0(h+ik)}\left[\left(1-\frac{ik}{h}\right)\cosh\omega+\frac{2h-ik-\frac{k^2}{h}}{\sqrt{4h^2-k^2}}\sinh\omega\right]\nonumber\\&-\frac{1}{2a_{out}^2}\left(1-\frac{ik}{h}\right)e^{2t_0h}\nonumber\\
\beta_{00}=&-\frac{2he^{4t_0h}}{a_{out}^2\sqrt{4h^2-k^2}}\sinh\omega
\end{align}
where $\omega=2t_0\sqrt{4h^2-k^2}$, $a_{out}$ is the scale 
factor in the {\it out} region and we have assumed 
that $a_{in}=1$. In Fig. 1 we show the corresponding production 
of temporal modes 
$n_0^{out}(k)=\vert \beta_{00}\vert^2
+\vert \beta_{\parallel 0}\vert^2$ for $h=0.29$ and $t_0=2$ 
(which approximately corresponds to the Hubble expansion rate 
used for the numerical computation). Notice that the approximate
analytical solution reproduces very accurately the numerical result
at small $k$. This is because, on that region, the production of longitudinal
modes is practically zero and for very large wavelenghts,
the production of temporal states is not sensitive to the particular
shape of the expansion rate.

These results show  that 
in a space-time configuration with asymptotic flat regions, an initial
state satisfying the weak Lorenz condition does not necessarily satisfy it
at a later time. In order to overcome this problem, 
it is possible to impose 
a more stringent gauge-fixing condition. Thus,  following \cite{Pfenning}
we can define 
the physical states $\vert \phi\rangle$ as those such that 
$\nabla_\mu A^{\mu (+)}\vert \phi\rangle =0$, 
$\forall \eta$. Although this is a perfectly consistent solution, notice that 
the separation in positive and negative frequency parts depends on 
the space-time geometry and therefore, the determination of the physical states 
requires a previous knowledge of 
the geometry of the universe at all times.

Another possible way out would be to modify the standard Gupta-Bleuler formalism
by including ghosts fields as done in non-abelian gauge theories \cite{Kugo}.
With that purpose, the action of the theory (\ref{actionF}) can be modified by including
the ghost term (see \cite{Adler}):
\begin{eqnarray}
S_{g}=\int d^4x\sqrt{g}\, g^{\mu\nu}\,\partial_\mu \bar c\, \partial_\nu c
\label{ghost}
\end{eqnarray} 
where $c$ are the complex scalar ghost fields. It is a well-known result 
\cite{Adler,BO} that by choosing appropriate boundary conditions  for the
electromagnetic and ghosts Green's functions, it is possible to get  
$\langle \phi\vert T_{\mu\nu}^\xi+T_{\mu\nu}^g\vert \phi\rangle=0$, where
$T_{\mu\nu}^\xi$ and $T_{\mu\nu}^g$ denote the contribution to the  energy-momentum
tensor from the $\xi$ term in (\ref{actionF}) and from the ghost term (\ref{ghost})
respectively.
Notice that a choice of boundary conditions in the Green's functions 
corresponds to a choice of  vacuum state. Therefore, also in this case  
an a priori knowledge of the future behaviour of the universe
geometry is required in order to determine the physical states.

In this work we follow a different approach in order to deal with the 
difficulties found in the Gupta-Bleuler formalism  and  we will explore
the possibility of quantizing  electromagnetism in an expanding
universe without imposing any subsidiary condition.

\section{Quantization without the Lorenz condition}

In the previous section it has been shown that 
although the Lorenz gauge-fixing conditions can be imposed in the covariant
formalism, this cannot be done in a straightforward way.   
These difficulties could be suggesting  some more fundamental obstacle 
in the formulation of an electromagnetic gauge invariant theory in an expanding universe.  
As a matter of fact, 
electromagnetic models which break gauge invariance on cosmological 
scales have been widely considered in the context of generation
of primordial magnetic fields (see, for instance, \cite{Turner}).

Let us then explore the possibility that the fundamental 
theory of electromagnetism is not given by the gauge invariant 
 action (\ref{action}), but by the gauge non-invariant action (\ref{actionF}). 
Notice that
although (\ref{actionF}) is not invariant under general gauge transformations, 
it respects
the invariance under residual ones and, as shown below, in Minkowski 
space-time, the theory
is completely equivalent to standard QED.    
Since  the fundamental electromagnetic 
theory is assumed non-invariant under arbitrary gauge transformations, then
there is no need to impose the Lorenz constraint in the 
quantization procedure. Therefore,  having removed one constraint, 
the theory  contains one additional degree of freedom. Thus, 
the general solution for the modified equations (\ref{EMeqexp})
can be written as:
\begin{eqnarray}
\A_\mu=\A_\mu^{(1)}+\A_\mu^{ (2)}+\A_\mu^{(s)}+\partial_\mu \theta
\end{eqnarray}
where $\A_\mu^{(i)}$ with $i=1,2$ are the two transverse modes of
the massless photon, $\A_\mu^{(s)}$ is the new scalar state, which
is the mode that would have been eliminated if we had imposed the
Lorenz condition and, finally, $\partial_\mu \theta$ is a purely
residual gauge mode, which can be eliminated by means of a
residual gauge transformation in the asymptotically free regions, 
in a completely analogous way to the elimination of the $A_0$
component in the Coulomb quantization.  The fact that
Maxwell's electromagnetism could contain an additional scalar
mode decoupled from electromagnetic currents, but with 
 non-vanishing  gravitational interactions, was already noticed 
in a different context in \cite{Deser}. 

In order to quantize the free theory, we perform the mode
expansion of the  field with the corresponding creation and
annihilation operators for the {\it three} physical states:
\begin{eqnarray}
\A_{\mu}=\int d^3\vec{k}\ \sum_{\lambda=1, 2,s}\left[{\bf
a}_\lambda(k)\A_{\mu k}^{(\lambda)} +{\bf
a}_\lambda^\dagger(k)\overline{\A_{\mu k}^{(\lambda)}}\, \right]
\end{eqnarray}
where the modes are required to be orthonormal with respect to the
scalar product (see for instance \cite{Pfenning}):
\begin{eqnarray}
\left(\A^{(\lambda)}_k,\A^{(\lambda')}_{k'}\right)&=&i\int_{\Sigma}d\Sigma_\mu\left[\,
\overline{\A_{\nu k}^{(\lambda) }}\;\Pi^{(\lambda')\mu\nu}_{k'}-
\overline{\Pi^{(\lambda)\mu\nu }_{k}}\;\A_{\nu k'}^{(\lambda')}\right]\nonumber\\
&=&\delta_{\lambda\lambda'}\delta^{(3)}(\vec k-\vec k'),\;\;\;\;\;
\lambda,\lambda'=1,2,s
\label{norm}
\end{eqnarray}
where $d\Sigma_\mu$ is the three-volume element of the Cauchy
hypersurfaces. In a Robertson-Walker metric in conformal time, it
reads $d\Sigma_\mu= a^4(\eta)(d^3x,0,0,0)$. The generalized
conjugate momenta are defined as:
\begin{eqnarray}
\Pi^{\mu\nu}=-(F^{\mu\nu}-\xi g^{\mu\nu}\nabla_\rho A^\rho)
\end{eqnarray}
Notice that
the three modes can be chosen to have positive normalization.
The equal-time commutation relations:
\begin{eqnarray}
\left[\A_\mu(\eta,\vec x),\A_\nu(\eta,\vec x\,')\right]=
\left[\Pi^{0\mu}(\eta,\vec x),\Pi^{0\nu}(\eta,\vec x,')\right]=0
\end{eqnarray}
and
\begin{eqnarray}
\left[\A_\mu(\eta,\vec x),\Pi^{0\nu}(\eta,\vec x\,')\right]=
i\frac{\delta_\mu^{\;\nu}}{\sqrt{g}}\delta^{(3)}(\vec x-\vec x\,')
\end{eqnarray}
can be seen to imply  the canonical commutation relations:
\begin{eqnarray}
\left[{\bf a}_\lambda(\vec{k}),{\bf a}_{\lambda'}^\dagger(\vec{k'})\right]
=\delta_{\lambda\lambda'}\delta^{(3)}(\vec{k}-\vec{k'}),\;\;\;
\lambda,\lambda'=1,2,s
\end{eqnarray}
by means of the  normalization condition in (\ref{norm}).
Notice that the sign of the commutators is positive for the
three physical states, i.e. there are no negative norm states
in the theory, which in turn implies that there are no
negative energy states as we will see below in an
explicit example.

Since  $\nabla_\mu\A^{\mu}$ evolves as a minimally coupled scalar field,
as shown in
(\ref{minimal}), on sub-Hubble scales ($\vert k\eta\vert \gg 1$),
we find that
for arbitrary background evolution,
$\vert \nabla_\mu\A^{(s)\mu}_k\vert \propto a^{-1}$, i.e.
the field is suppressed by the universe expansion, thus 
effectively recovering the Lorenz condition on small scales. Notice that this is  
a consequence of the cosmological evolution,  not being imposed 
 as a boundary condition as in the flat space-time case.

On the other hand,
 on super-Hubble scales ($\vert k\eta\vert \ll 1$),
$\vert\nabla_\mu\A^{(s)\mu}_k\vert= const.$ which, as shown in
\cite{EM}, implies that the field contributes as a cosmological
constant in (\ref{actionF}). Indeed, the energy-momentum tensor 
derived from (\ref{actionF}) reads:
\begin{eqnarray}
T_{\mu\nu}&=&-F_{\mu\alpha}F_\nu^{\;\;\alpha}
+\frac{1}{4}g_{\mu\nu}F_{\alpha\beta}F^{\alpha\beta}\nonumber\\
&+&\frac{\xi}{2}\left[g_{\mu\nu}\left[\left(\nabla_\alpha
A^\alpha\right)^2 +2A^\alpha\nabla_\alpha\left(\nabla_\beta
A^\beta\right)\right] -4A_{(\mu}\nabla_{\nu)}\left(\nabla_\alpha
A^\alpha\right)\right]
\end{eqnarray}
Notice that for the scalar electromagnetic mode in the 
super-Hubble limit, the contributions 
involving $F_{\mu\nu}$ vanish and only the piece proportional
to $\xi$ is relevant. 
 Thus, it can be easily seen that, since in this case
$\nabla_\alpha A^\alpha=constant$, the energy-momentum tensor is just given by:
\begin{eqnarray}
T_{\mu\nu}=\frac{\xi}{2} g_{\mu\nu}(\nabla_\alpha A^\alpha)^2
\end{eqnarray}
which is the energy-momentum tensor of a cosmological constant.  
Notice that,  as seen in (\ref{minimal}), the new scalar mode is a massless free field.  
This is one of the most relevant aspects of the present model
in which, unlike existing dark energy theories based on scalar fields, 
dark energy can be generated without including any potential term
or dimensional constant.

Since, as shown above, the field amplitude
remains frozen on super-Hubble scales and starts decaying once the
mode enters the horizon in the radiation or matter eras, the
effect of the $\xi$ term in (\ref{EMeqexp}) is completely
negligible on sub-Hubble scales, since the initial amplitude
generated during inflation is very small as we will show below. 
Thus, below 1.3 AU,
which is the largest distance scale at which electromagnetism has
been tested \cite{Nieto}, the modified Maxwell's equations
(\ref{EMeqexp}) are physically indistinguishable from the flat
space-time ones (\ref{Maxwell}).

Notice that in Minkowski space-time, the  theory (\ref{actionF})
is completely equivalent to standard QED. This is so because, although
non-gauge invariant, the corresponding effective action is equivalent to the 
standard BRS invariant effective action of QED. 
Thus, the effective action for
QED obtained from (\ref{Maxwell}) by the standard gauge-fixing procedure reads:
\begin{eqnarray}
e^{iW}=\int [dA][dc][d\bar c][d\psi][d\bar\psi] e^{i\int d^4x 
\left(-\frac{1}{4}F_{\mu\nu}F^{\mu\nu}+\frac{\xi}{2}
(\partial_\mu A^\mu)^2+\eta^{\mu\nu}\partial_\mu\bar c\, \partial_\nu c
+ {\cal L}_F\right)}
\end{eqnarray}
where ${\cal L}_F$ is the Lagrangian density of charged fermions. The $\xi$
term and the ghosts field appear in the Faddeev-Popov procedure when 
selecting an element of each gauge orbit.  
However, ghosts being decoupled from the electromagnetic currents can be integrated
out in flat space-time, so that up to an irrelevant normalization constant we find:
\begin{eqnarray}
e^{iW}=\int [dA][d\psi][d\bar\psi] 
e^{i\int d^4x \left(-\frac{1}{4}F_{\mu\nu}F^{\mu\nu}+\frac{\xi}{2}
(\partial_\mu A^\mu)^2+ {\cal L}_F\right)}
\end{eqnarray}
which is nothing but the effective action coming from 
the gauge non-invariant theory (\ref{actionF}) in flat space-time, in which
no  gauge-fixing procedure is required.

\section{An explicit example: quantization in de Sitter space-time}

Let us consider an explicit example which is
given by the  quantization in an inflationary de Sitter space-time with
$a(\eta)=-1/(H_I\eta)$. Here we will take $\xi=1/3$, although similar
results can be obtained for any $\xi>0$.  For this case, in which $\h'=\h^2$, we can obtain the following expression for the temporal component in terms of the longitudinal one from (\ref{coupledeq}):
\begin{equation}
\A_{0k}=\frac{-i}{2k(4k^2+3\h^2)}\left[2\frac{d^3\A_{\parallel k}}{d\eta^3}-\h\frac{d^2\A_{\parallel k}}{d\eta^2}+10k^2\frac{d\A_{\parallel k}}{d\eta}-5\h k^2\A_{\parallel k}\right].
\end{equation}
On the other hand, if we insert this relation into the equation for $\A_{\parallel k}$ we obtain a fourth-order equation for the longitudinal component which is completely decoupled from $\A_{0 k}$. This fourth-order equation can be expressed as follows:
\begin{equation}
\left[\frac{d^2}{d\eta^2}-4\frac{2k^2+3\h^2}{4k^2+3\h^2}\h\frac{d}{d\eta}+
\frac{4k^4+6\h^4-9k^2\h^2}{4k^2+3\h^2}\right]\left[\frac{d^2}{d\eta^2}+2\h\frac{d}{d\eta}+k^2\right]\A_{\parallel k}=0\label{eqfac},
\end{equation}
where we have performed a factorization in such a way that the second bracket is nothing but the operator determining the evolution of a free scalar field, i.e., $\Box\theta=0$.  Now, we shall use the notation ${\mathcal{F}}({\mathcal{G}} (\A_{\parallel k}))=0$, where ${\mathcal F}$ and ${\mathcal {G}}$ are the operators given in the first and second brackets in (\ref{eqfac}) respectively. In particular, ${\mathcal {G} }=\Box$ as we have already said. Thus, in order to solve the equation we have to obtain the kernel of ${\mathcal {F}}$, whose solutions will be denoted by $S$, and, then, solve the equation for a free scalar field with $S$ as an external source: ${\mathcal {G}}(\A_{\parallel k})=S$. This method has the advantage of allowing to identify the pure residual gauge mode as the homogeneous solution of this last equation. The explicit solution for the normalized scalar state is:
\begin{eqnarray}
\A_{0k}^{(s)}&=&-\frac{1}{(2\pi)^{3/2}}\frac{i}{\sqrt{2k}}
\left\{k\eta e^{-ik\eta}\right.\nonumber\\
&+&\left.\frac{1}{k\eta}\left[\frac{1}{2}(1+ ik\eta)
e^{-ik\eta}-k^2\eta^2e^{ik\eta}E_1(2ik\eta)\right]\right\}e^{i\vec k \vec x}\nonumber\\
\nonumber\\
\A_{\parallel k}^{(s)}&=&\frac{1}{(2\pi)^{3/2}}\frac{1}{\sqrt{2k}}
\left\{(1+ik\eta)e^{-ik\eta}\right.\nonumber\\
&-&\left.\left[\frac{3}{2}e^{-ik\eta}+(1-ik\eta)e^{ik\eta}
E_1(2ik\eta)\right]\right\}e^{i\vec k \vec x}\label{scalar}
\end{eqnarray}
 where $E_1(x)=\int_1^\infty e^{-tx}/tdt$ is the exponential
integral function and we have fixed the integration constants so that the mode is canonically normalized according to (\ref{norm}) . Using this solution, we  find:
\begin{eqnarray}
\nabla_\mu\A^{(s)\mu}_k=-\frac{a^{-2}(\eta)}{(2\pi)^{3/2}}
\frac{ik}{\sqrt{2k}}\frac{3}{2}\frac{(1+ik\eta)}
{k^2\eta^2}e^{-ik\eta+i\vec k \vec x}
\end{eqnarray}
so that the field is suppressed in the sub-Hubble limit
as  $\nabla_\mu\A^{(s)\mu}_k\sim \Od((k\eta)^{-2})$.

On the other hand, from the energy density given by $\rho_A= T^0_{\;\;0}$,
we obtain in the sub-Hubble limit the corresponding Hamiltonian, which is
given by:
\begin{eqnarray}
H=\frac{1}{2}\int \frac{d^3\vec
k}{a^4(\eta)}k\sum_{\lambda=1,2,s}\left[ {\bf
a}_{\lambda}^{\dagger}(\vec k){\bf a}_{\lambda}(\vec k) +{\bf
a}_{\lambda}(\vec k){\bf a}_{\lambda}^{\dagger}(\vec k) \right].
\end{eqnarray}
We see that the theory does not contain negative energy states (ghosts).
In fact, as shown in \cite{EM,VT}, the theory does not exhibit
either local gravity inconsistencies or classical instabilities.

Finally, from (\ref{scalar}) it is possible to obtain the
dispersion of the effective cosmological constant during
inflation:
\begin{eqnarray}
\langle 0\vert(\nabla_\mu\A^{\mu})^2\vert 0 \rangle=\int\frac{dk}{k}P_A(k)
\end{eqnarray}
with $P_A(k)=4\pi k^3\vert\nabla_\mu\A^{(s)\mu}_k\vert^2 $. In the
super-Hubble limit, we obtain for the power-spectrum:
\begin{eqnarray}
P_A(k)=\frac{9H_I^4}{16\pi^2},
\end{eqnarray}
in agreement with \cite{EM}. Notice that this
result implies that $\rho_A\sim (H_I)^4$. The measured value of
the cosmological constant then requires $H_I\sim 10^{-3}$ eV,
which corresponds to an inflationary scale  $M_I\sim 1$ TeV.
Thus we see that the cosmological constant scale can be naturally
explained in terms of physics at the electroweak scale.

Once we have computed the initial amplitude of the 
electromagnetic fluctuations, we can estimate the
magnitude of the corrections induced by the new
terms in the modified Maxwell's equations. Notice that
(\ref{EMeqexp}) can be rewritten as the ordinary Maxwell's equation
with an additional conserved current source:
\begin{eqnarray}
\nabla_\nu F^{\mu\nu}=J^\mu+J^\mu_\xi
\label{EMeqexp2}
\end{eqnarray}
where $J^\mu_\xi=-\xi\nabla^\mu(\nabla_\nu A^\nu)$  
satisfies $\nabla_\mu J^\mu_\xi=0$ by virtue of (\ref{minimal}).
For sub-Hubble modes below $1.3$ A.U., which is the region on which
electromagnetism has been tested, consistency requires the 
new current to be negligible. It is possible to estimate the 
present value of such a current by noticing that a Fourier mode
$k$ entered the Hubble radius in the radiation dominated era when the 
scale factor was $a_{in}\simeq H_0\, a_{eq}^{1/2}/k$, with $a_{eq}$ the scale
factor at matter-radiation equality and where we have taken that today $a_0=1$.
Therefore, since the initial amplitude $\nabla_\mu A^\mu\simeq H_I^2$ 
remains constant until horizon reentry, today we obtain for the effective 
electric charge density created by the mentioned modes:
\begin{eqnarray}
J^0_\xi\sim k \,H_I^2 \,a_{in}\sim H_0\, H_I^2\, a_{eq}^{1/2}\sim 10^{-41}\,
 \mbox{eV}^3
\end{eqnarray}
which is independent of $k$ and completely negligible since it 
roughly corresponds to the charge density of one electron in the 
volume of the Earth.

\section{Conclusions}

In this work we have considered the difficulties 
in the application of the covariant quantization method for electromagnetic
fields in an expanding universe.  Although the 
Lorenz gauge-fixing condition can in principle be formally applied 
also in cosmological contexts, we have explored the alternative possibility 
of quantization without subsidiary 
conditions, and therefore, without the introduction of ghost fields.

In this approach the theory is  invariant only under
 residual gauge transformations and a new scalar mode appears for the 
electromagnetic field.  Despite these facts, 
standard QED is recovered in the flat space-time limit
as the new scalar state is
completely decoupled from the conserved electromagnetic currents 
(only transverse photons couple to them \cite{Pokorski}).
 The residual
gauge symmetry of the theory ensures that the negative norm state can
be eliminated and that the energy is positive as  shown in an
explicit example. 

This quantization procedure is found to have interesting consequences for dark energy. 
Thus, the energy
density of the new mode on super-Hubble scales (which is essentially the temporal
part of the electromagnetic field) is seen to behave as a cosmological constant
irrespectively of the expansion rate of the universe.
The new mode can be generated during inflation, in a similar way to 
the inflaton fluctuations and this fact allows to establish a link between the scale of
inflation and the value of the cosmological constant.

\vspace{0.2cm}

{\em Acknowledgments:}
 This work has been  supported by
Ministerio de Ciencia e Innovaci\'on (Spain) project numbers
FIS 2008-01323 and FPA
2008-00592, UCM-Santander PR34/07-15875, CAM/UCM 910309 and
MEC grant BES-2006-12059.

\end{document}